\documentclass{sf2a-conf2019}
\usepackage{graphicx}
\usepackage{hyperref}
\usepackage[]{natbib}  
\usepackage{epstopdf}

\def\BibTeX{{\rm B\kern-.05em{\sc i\kern-.025em b}\kern-.08em
    T\kern-.1667em\lower.7ex\hbox{E}\kern-.125emX}}
\bibpunct{(}{)}{;}{a}{}{,}  

\begin{document}

\TitreGlobal{SF2A 2019}


\title{Are Milky-Way dwarf-spheroidal galaxies dark-matter free?}

\runningtitle{No dark matter in MW dSphs?}

\author{F. Hammer}\address{GEPI, Observatoire de Paris, Universit\'e PSL, CNRS, Place Jules Janssen 92195, Meudon, France}

\author{Y. B. Yang$^1$}
\author{J. L. Wang$^1$}\address{NAOC, Chinese Academy of Sciences, A20 Datun Road, 100012 Beijing, PR China.}
\author{F.Arenou$^1$}
\author{C. Babusiaux}\address{Universit\'e de Grenoble-Alpes, CNRS, IPAG, F-38000 Grenoble, France}
\author{M. Puech$^1$}
\author{H. Flores$^1$}




\setcounter{page}{237}


\maketitle


\begin{abstract}
We have found that the high velocity dispersions of dwarf spheroidal galaxies (dSphs) can be well explained by Milky Way (MW) tidal shocks, which reproduce precisely the gravitational acceleration previously attributed to dark matter (DM). Here we summarize the main results of \citet{Hammer2019} who studied the main scaling relations of dSphs and show how dark-matter free galaxies in departure from equilibrium reproduce them well, while they appear to be challenging for the DM model. These results are consistent with our most recent knowledge about dSph past histories, including their orbits, their past star formation history and their progenitors, which are likely tiny dwarf irregular galaxies. 
\end{abstract}



\section{Introduction}
  DSphs (including ultra-faint dwarfs, UFDs) in the Milky Way halo are by far the smallest galaxies that can be detected and studied. They are believed to contain large amounts of DM, which fraction is generally assumed to increase with decreasing luminosity or stellar mass \citep{Strigari2008,Walker2009,Wolf2010}. The existence of DM has been widely accepted since the discovery of the HI-extended and flat rotation curves in giant spirals \citep{Bosma1978}. The extent of the DM paradigm towards the dwarf galaxy regime has been initiated by \citet{Aaronson1983} on the basis of the too large velocity dispersion of the Draco stars. This pioneering result was only based on 3 stars in Draco, though it has been confirmed by major works that have identified new dwarfs, measured their distances, and performed deep photometry and high resolution spectroscopy of their individual stars. During the last 35 years these long term works have provided robust measurements for several tens of MW dSphs \citep[and references therein]{Munoz2018,Fritz2018}.\\
The scaling relations between the visible luminosity, the half light radius, the velocity dispersion and the MW distance can be established for 24 dSphs possessing sufficiently robust measurements. Three dSphs (Sagittarius, Crater II and Bootes I) are clearly outliers in these relations, which leads to a sample of 21 dSphs. 
Analyzing these data, \cite{Hammer2019} demonstrated that the MW gravitation through tidal shocks can fully account for the dSph kinematics. DM estimates \citep{Walker2009,Wolf2010} are based on only the projected mass density along the line-of-sight. \citet[see their Fig. 1]{Hammer2019} found that this quantity is highly anti-correlated with the MW distance, a property that cannot be reproduced by DM-dominated models. It is however naturally expected if dSphs are tidally shocked during their first passage into the MW halo. 

\section{DSph progenitors:  A first infall of gas-rich dwarfs in the MW?}
\begin{figure}[ht!]
\centering
\includegraphics[width=0.45\textwidth,clip]{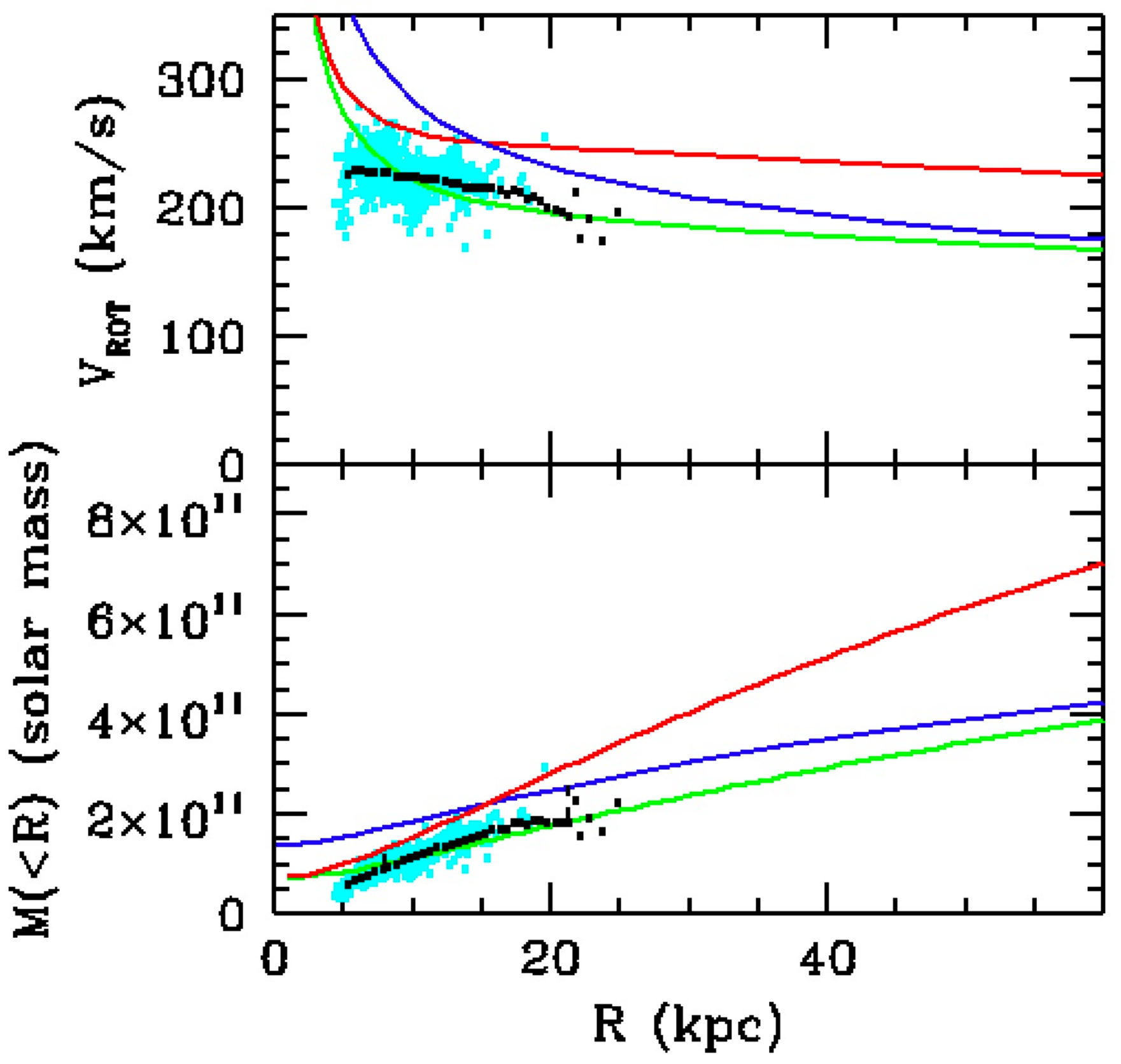}%
 \includegraphics[width=0.45\textwidth,clip]{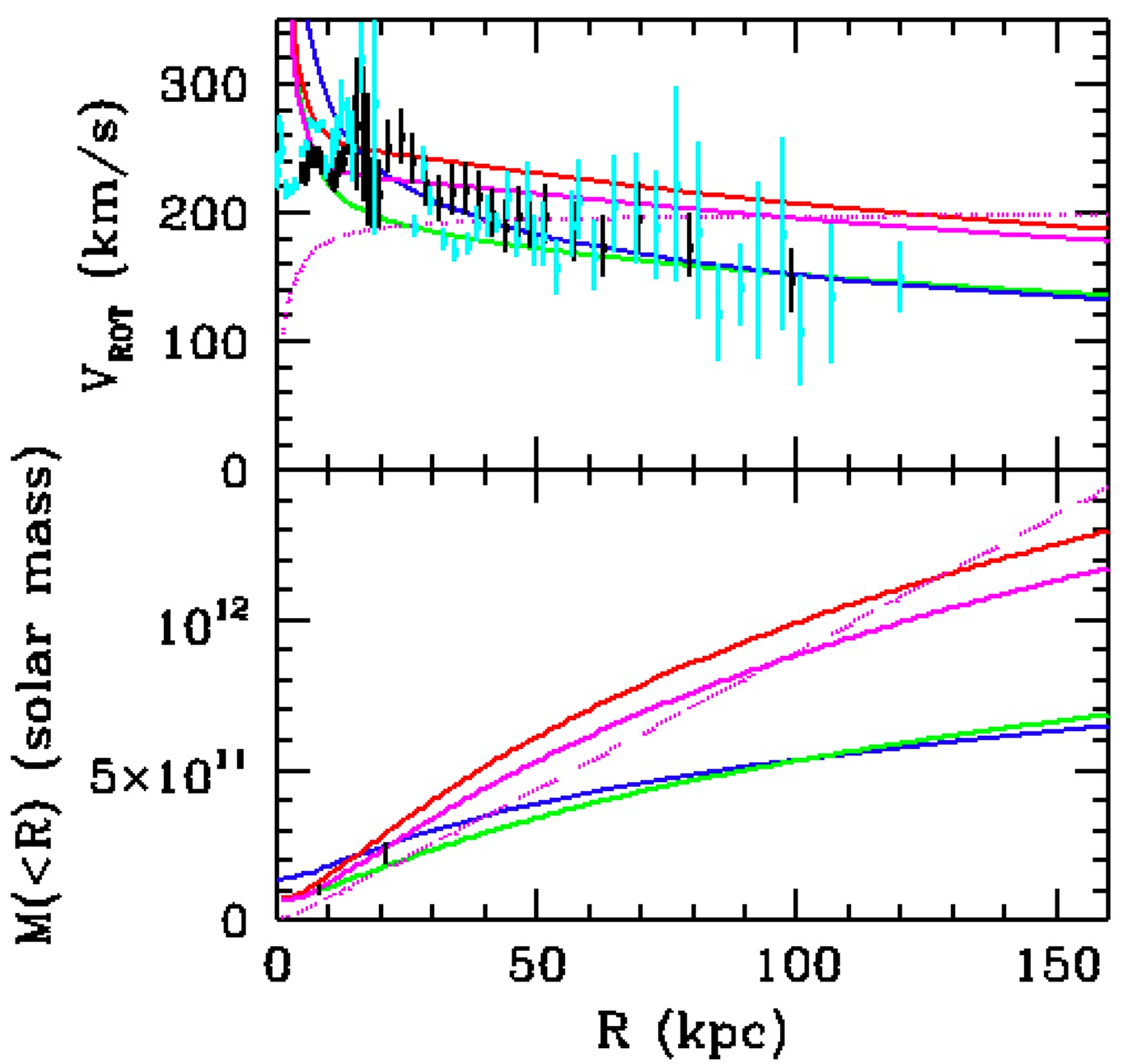}      
\caption{{\bf Top left:} Rotation curve of the MW. Black and cyan points represent the new determinations by \citet{Eilers2019} using massive stars, and by \citet{Mroz2019} using cepheids, respectively.  The green line shows the  \citet{Bovy2015} model, the red line represents the \citet{Fritz2018} model for which they have multiplied the halo mass by 2, and the blue line shows the model of \citealt{Sofue2015}.  {\bf Bottom left:} MW mass profile, same symbols as in the top left panel. {\bf Top right:} Extended rotation curve of the MW. Black and cyan points with error-bars represent compilations from \citet{Huang2016} and \citet{Bhattacharjee2014}, respectively.  As in the left panels, the green, red and blue lines show the \citet{Bovy2015} model, the \citet{Fritz2018} model for which they have multiplied the halo mass by 2, and the \citet{Sofue2015} model, respectively. The magenta (full and dotted) lines show the models of \citealt{McMillan2017} and \citealt{Irrgang2013}, respectively.  {\bf Bottom right:} MW mass profile, same symbols as in the top right panel. }
   \label{author1:fig1}
\end{figure}

GAIA DR2 is revolutionizing our knowledge of the MW dSphs orbits in a two-fold way: 
\begin{enumerate}
\item It has considerably improved our knowledge of the MW mass distribution up to 20-50 kpc, by establishing a more accurate rotation curve \citep{Eilers2019,Mroz2019}, and by providing better constraints on the Globular Cluster motions \citep{Eadie2019} and on the estimates of the escape velocity \citep{Deason2019}. These studies provide MW masses ranging from 0.7 to 1 $\times 10^{12} M_{\odot}$ (see however \citealt{Grand2019} for a slightly higher value), and all of them seem to exclude larger masses. Fig.~\ref{author1:fig1} shows the different rotation curves and mass distributions of the MW, including the most recent ones \citep{Eilers2019,Mroz2019} and the (former) high-mass models \citep[with total mass from 1.37 to more than 1.9 $\times 10^{12} M_{\odot}$]{McMillan2017,Irrgang2013} adopted by \citet{Gaia2018} during the release of dSph proper motions. Top panels of Fig.~\ref{author1:fig1} show that high mass models lead to velocities (red and magenta lines) much higher than that observed (points).
\item The above determined MW mass range allows to calculate accurate orbits that are generally consistent with a first infall for most dSphs. For example 2/3 of them have eccentricities in excess of 0.66 (such as the LMC) and half of them with apocenter larger than 300 kpc \citep{Fritz2018} when adopting the \citet{Bovy2015} MW mass model that reproduces its kinematics.\\
\end{enumerate}

 

DSph progenitors are likely gas-stripped dwarfs due to the ram-pressure caused by the MW halo gas as it has been proposed by \citet{Mayer2001}. Observations support this scenario because all dwarfs (but the Clouds) are gas-rich beyond 300 kpc and gas poor within 300 kpc \citep{Grcevich2009}. The gas removal by ram-pressure induces a lack of gravity implying that stars are then leaving the system following a spherical geometry. Such a geometry ensures the dominance of tidal shocks over tidal stripping \citep{Binney2008} explaining the absence of tidal features in most dSphs \citep{Hammer2019}.  MW tidal shocks increase the square of the velocity dispersion by $\sigma_{\rm MWshocks}^2=  \sqrt{2} \: \alpha_{\rm MW}\;  g_{\rm MW} \: r_{\rm half} \:$ where $g_{\rm MW}$ is the MW gravitational acceleration and $\alpha_{\rm MW} = 1 - \partial log(M_{\rm MW})/\partial log(D_{\rm MW})$  \citep{Hammer2018}. This property reproduces quite precisely the observed dSph velocity dispersions as well as the fundamental relationships established from the observations (see Figs. 1-3 and 5-7 in \citealt{Hammer2019}). \\

The role of the gas during the process is essential. First, it would be very unlikely that dSphs progenitors were without gas since such objects are extremely rare in the field. Second, DM-devoid models made by \citet{Piatek1995} assumed gas-free progenitors, which implies a strong dominance of tidal stripping. There are similar models \citep{Kroupa1997,Klessen1998,Iorio2019} that also assumed multiple orbits furthermore limiting the possibility that tidal shocks affect the dSph velocity dispersions. 

\section{Discussion and Conclusion}

\begin{figure}[ht!]
 \centering
 \includegraphics[width=0.65\textwidth,clip]{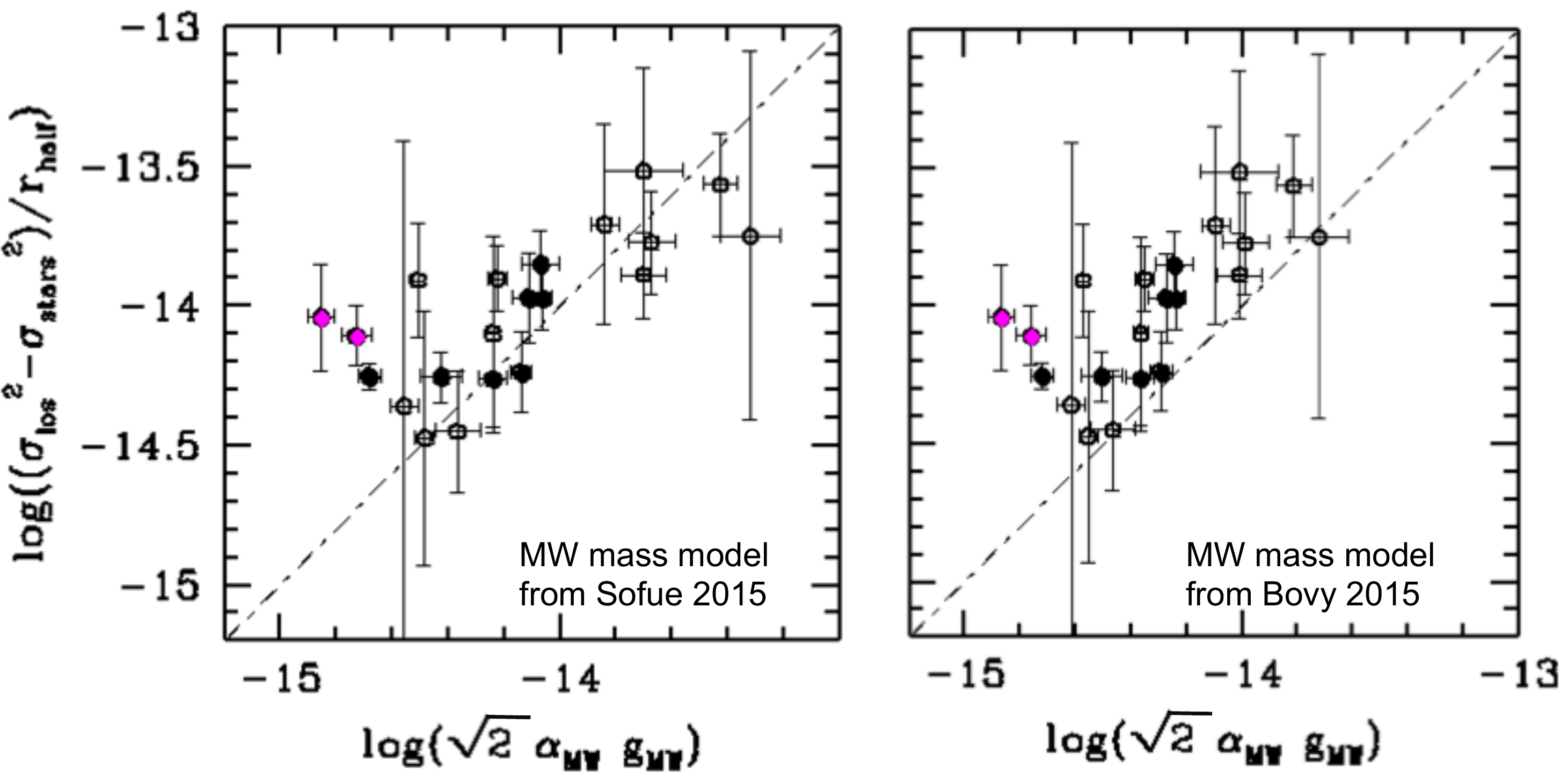}      
  \caption{Self-gravity acceleration $a_{DM}= ( \sigma_{\rm los}^2 - \sigma_{\rm stars}^2) \times r_{\rm half}^{-1}$ derived from DM estimates in dSphs compared (in logarithmic scale) to that due to MW tidal shocks, $a_{\rm MWshocks}= \sqrt{2} \: \alpha_{\rm MW}\;  g_{\rm MW} \:$. The left and the right panel show the relation for an adopted MW mass profile from \cite{Sofue2015} and from \cite{Bovy2015}, respectively. Full (open) dots represent classic (non-classic) dSphs, respectively. Leo I and II are identified by points with magenta color since they do not fully obey the impulse approximation (see \citealt{Hammer2019}). Their location in the Figure can be well explained if they are affected by tidal stripping in addition to tidal shocks.}
   \label{Hammer:fig2}
\end{figure}

The DM content of dSphs derived by \citet{Walker2009} and by \citet{Wolf2010} comes from the measurement of the dSph-DM self-gravity acceleration projected along the line of sight, which is $a_{\rm DM}= G M_{\rm DM} \times r_{\rm half}^{-2} = ( \sigma_{\rm los}^2 - \sigma_{\rm stars}^2) \times r_{\rm half}^{-1}$. \citet{Hammer2019} showed that over more than a decade, $a_{\rm DM}$ matches very well with the acceleration caused by MW tidal shocks on DM-free dSphs, which is $a_{\rm MWshocks}= \sqrt{2} \: \alpha_{\rm MW}\;  g_{\rm MW} \:$ (see Fig.~\ref{Hammer:fig2}). Why would the acceleration caused by the DM be precisely what it is expected from MW tidal shocks on DM-free dSphs? Why do MW tidal shocks predict that the DM mass-to-light ratio of Segue 1 is several thousand, while that of Fornax is around ten? The acceleration caused by the DM ($a_{\rm DM}$) also strongly anti-correlates with the dSph distance from the MW \citep[see their Fig. 1]{Hammer2019}. The probability that this is just a coincidence is only 3$\times 10^{-4}$, which can be conservatively considered as the chance that DM impacts the kinematics of dSphs.  \\

We are aware that the above could seriously affect the paradigm of DM in MW dSphs, and then have an impact on the cosmological models, except if MW dSphs are not representative of the dwarf regime. It is not unexpected that these results would be met with some skepticism. Could this be contradicted by other properties, e.g., of their progenitors that are likely dwarf Irregulars (dIrrs)? The DM content of dIrrs that share a similar stellar mass range than dSphs is still not well constrained. The most massive dSph (e.g., Fornax and maybe Sculptor) progenitors can be found, e.g., in the smallest galaxies of the sample from \citet{Lelli2016}, which includes the best studied rotation curves over a large mass range. The DM content of these small galaxies that are all dIrrs has been derived from their rotation curves and varies from none to large values, in particular within a radius similar to the half light radius that is adopted for sampling dSph dynamical properties \citep{Walker2009,Wolf2010,Hammer2019}. It appears very hard to assess rotation curves and velocity amplitudes in dwarf galaxies that are too irregular. 
Trying to identify possible progenitors of smaller dSphs (e.g., UFDs) that constitute the bulk of dSphs leads one to consider extremely small dIrrs, for which establishing their rotation curves can not be seriously attempted \citep{McNichols2016,Oh2015,Ott2012}. This is because these tiny objects are far from being represented by a thin disk geometry and also because their velocity amplitudes are similar to that of their dispersion, the latter being mostly associated to star formation and turbulence \citep{Stilp2013}.\\

Star formation histories of dSphs have been well studied especially for the most massive ones \citep{Weisz2014}.  For Fornax \citep{de Boer2013} it is consistent with a recent gas removal by ram-pressure and then with the tidal shock scenario. While this also applies to Carina, Leo I and perhaps to Leo II, past histories of Sculptor, Sextans, UMi and Draco \citep{Weisz2014} are perhaps more problematic.  Why did the star formation in the Sculptor progenitor stop about 5 Gyr ago \citep{de Boer2012} if it was still in isolation at a later time? Only a full hydrodynamical simulation with a well determined orbital history for Sculptor would help us to verify a potential inconsistency. Interestingly,  recent simulations \citep{Garrison-Kimmel2019} have shown that gas-rich dwarfs with Sculptor stellar mass and in isolation may have similar star formation histories than Sculptor, and this also applies to galaxies with smaller masses that form the bulk of the MW dSphs. \\
The above may lead to a significant change of paradigm in our understanding of the MW dSphs, which could impact the determination of the lower end of the galaxy mass function. Next steps will be to verify if this is consistent with other dSphs of the MW, which total mass can be robustly determined without extrapolations.

\end{document}